# Low-Cost HEM with Arduino and Zigbee Technologies in the Energy Sector in Colombia

Zurisaddai de la Cruz Severiche Maury [1], Ana Fernandez Vilas [2] and Rebeca Diaz Redondo [2,*]

[1] Electronics Research and Innovation Group (GINELECT), Universidad de Sucre, Sincelejo 700001, Colombia; severichemaury@gmail.com

[2] I & C Lab. AtlanTTic Research Centre, University of Vigo, 36310 Vigo, Spain; avilas@det.uvigo.es

* Correspondence: avilas@det.uvigo.es

**Abstract:** Since no solutions have been proposed in Colombia that seek to reduce the consumption of electricity at the residential level, this paper describes the design and implementation of a simple prototype of a low-cost home energy management system (HEMS). The objective of this platform is to monitor the energy consumption of typical household devices so that users can access the consumption of each device separately and then establish the strategy that allows them to reduce energy consumption at home. In order to demonstrate that our system is viable, the system has been evaluated by measuring weekly energy consumption with the on-line and off-line HEMS using a test bench with typical household devices in a Sincelejo typical household. The evaluation has shown that with the installation of this HEMS, consumption is reduced by 27%. This shows that it is possible to achieve a good reduction percentage with a low-cost system.

## 1. Introduction

A smart grid is an energy delivery system that moves from a centrally controlled system, like the ones we currently have, to a consumer-driven approach, i.e., an iterative system relying on bi-directional communication to adapt and tune the delivery of energy in the real-time market. A smart grid includes a broad range of sophisticated sensors that constantly assess the state of the grid and the electrical power demand and availability, with the aim of optimizing the energy supply. In the future, the power grid will evolve into a cyber-physical system where smart devices will allow advanced monitoring and control. In this context, along with the smart home, consumers play more active roles in the management of energy consumption, and Home Energy Management Systems (HEMSs) emerge as hardware and software technology to monitor and to provide feedback on energy consumption at home and, so, to support effective habits from the point of view of energy saving. HEMSs are a growing sector in the modern era of the smart grid and smart homes.

Specifically, in Colombia, people have become increasingly dependent on energy, so that consumption is increasing year after year. Considering that the end user generates a notable amount of total energy consumption, and that the residential sector makes a considerable contribution, the DPN of the Colombian government (National Planning Department; by its acronym in Spanish) has included this sector in the energy efficiency goals in the country, for which reason HEMS technologies have been considered energy efficiency applications [1].

Those HEMSs have emerged as a solution that allows the residential sector to manage the energy consumption more efficiently. However, according to [2], energy management technologies for residential users have not yet been introduced in Colombia; [3] suggests that economic drawbacks are the main barrier to HEMS adoption. In addition to this, not only are there few companies in Colombia that focus on the objective of saving energy, but these companies offer devices that are not connected to a network, which would allow the control and management of energy consumption in real time. In addition, home automation systems are purchased by the high-income population because the outright price and cost of implementation are expensive for the rest of the population. The Colombian strategy of reducing energy consumption in residential areas needs an affordable mechanism to support the low-income population at home.

In this special context, this paper reports the collaborative work between the I&C Lab of the University of Vigo (Spain) and the University of Sucre (Colombia) to develop a low-cost solution and system to monitor the consumption of electricity at home in the pursuit of the reduction in energy consumption at the residential level. This system is applied to the Colombian context through the SAEH project of the SENA (National Learning Service; by its acronym in Spanish).

In our previous works [4,5], we have explored the needs and restrictions in Colombian households according to social strata, and we have reviewed the available technologies to present a low-cost and viable architectural proposal in the analyzed context. In this article, we introduce an affordable HEMS and describe its design, its implementation, and the results of pilots. This applied research evaluates the performance of a prototype implementation of a low-cost HEMS to show that HEMSs represent a viable alternative for energy saving according to the population analysis that allowed us to identify technical requirements to obtain greater energy savings within the dwellings. This prototype implementation can be applied in geographical areas with similar socio-economic and demographic characteristics.

The rest of the paper is organized as follows. Section 2 summarizes the related work, Section 3 describes materials and methods, Section 4 includes the energy demands in a Colombian Community, Section 5 shows the design of the proposed architecture, and the technological and implementation details are found in Section 6. Finally, validation and results are explained in Sections 7 and 8, and our conclusions and future work suggestions are found in Section 9.

## 2. Related Work

HEMSs, together with control strategies, optimize energy consumption. Several authors have worked with HEMSs with good results, which shows that the use of HEMSs is efficient in reducing energy consumption and allows saving money by reducing the cost of electricity bills. Table 1 presents a comparison between several HEMSs developed and ours, so that it can be seen how each one manages to achieve a reduction in energy consumption at home.

**Table 1.** Comparative study of HEM systems in the last 5 years.

| Ref/Com. Tech | Architectural Overview | Operation | Consumption Reduction |
|---|---|---|---|
| [6] GSM | The central component of the system is an Arduino UNO board, of which 20 pins are used for the interface of sensors and actuators. To detect the presence of the user, three PIC HC-SR501 modules are connected to the Arduino. Changes in light intensity are detected through photoresistors (LDR). | In this configuration, the main function of the system is to control the outlets to avoid wasting standby power. To minimize losses due to user negligence, their secondary role is to control the lights and, in certain situations, to turn them off automatically. | Standby power: 52.77%. Total energy consumption: 5.8% |
| [7] Power line | It uses microcontroller-based modules to make energy measurements. The HEM software | Measure power consumption in real time and monitor the supply of all connected home devices to optimize total power | 10% |

| | | consumption based on power consumption history to control devices. | |
|---|---|---|---|
| [8] Power line—Wi-Fi | It proposes a two-level communication system connecting the microgrid system, implemented in MATLAB, to the server in the cloud. | It presents a scheduling scheme for Internet of Energy (IdT)-based real-time home energy management systems. The scheme is a multi-agent method that considers two main purposes, including user satisfaction and the cost of energy consumption. The scheme is designed in a microgrid environment. | 21.3% y 31.335% |
| [9] RFID | The HEMS consists of an array of sensors, a high-end microcontroller, and relay banks. The sensor array consists of an RFID reader, temperature, humidity, and current sensors. | The system allows users to remotely control devices and generate invoices online through an easy-to-use mobile user interface application. | NA |
| [10] Wi-Fi | It is based on the Arduino platform; it uses a PZEM-004 to measure the electrical parameters and a DTH11 to measure the environmental parameters. It uses a smartphone user interface based on the Android system to display the collected data that is transmitted wirelessly through an esp8266 interface. | In this system, a transmitting and receiving node supports data communication in the HEMS configuration, which forms HANs (Home Area Networks). The real-time data collected at the central node can be used to prioritize and schedule the devices. You can estimate different power parameters and accumulate real-time energy consumption data from individual appliances. | There is a dramatic improvement in the efficiency of energy use and a significant decrease in the electricity bill. |
| [11] Data bus | It has a predictive control core (MPC), which generates control decisions for individual devices based on the information generated by a set of module inputs. | It presents a user-centered home energy management system that can help optimize the operation of a home to meet user needs, achieve energy efficiency and proportionally save utility costs, and offer a service network reliably based on public service signals. | 7.6% |
| [12] Z-Wave—Wi-Fi | It includes a household energy meter that measures total household demand, plug-in smart switches used as on/off modules, and relay switches connected to the device interfaces. The controller uses a Raspberry Pi with a ZWave transceiver from the RaZberry platform. Uses a Wi-Fi modem to allow the controller to communicate with the user interface over the Internet. The user interface was implemented as an Android application and a cloud server. | It reduces the maximum demand of a home by keeping the total demand of the home below a limit set by the user. The system uses the AS/NZS 4755 standard for demand response compliant devices. Combines this with plug-in modules connected to any mains powered device and a simple battery power storage system. | 48% |
| [13] ZigBee | It is a common standard house with all the necessary appliances, which is equipped with smart plugs designed | The local smart controller controls each of the household appliances by swapping them between battery and grid or | 15% |

| | | |
|---|---|---|
| and manufactured by the developers, a local controller and a photovoltaic system as a supplementary source. In general, the HEMS consists of a local controller and smart plugs. | postponing it to the other time interval to optimize the price of electricity, according to the grid price information, the convenience of residents and loading priority. Smart plugs provide an interface between the local controller and selected devices; these can monitor and measure the power consumption of the related electrical appliance and communicate with the local controller to determine and control the electrical appliance. | |
| [5] Zigbee | It consists of sensor-actuator nodes of four different types (temperature and humidity, luminosity, presence, and energy consumption), a coordinating node, and a control center. Mobile devices can access the system over the Internet through a gateway | The system consists of a sensor board that is connected to an Arduino development platform, this board consists of a sensor of environmental measurements or energy consumption and a Zigbee module that will allow the transmission of the data to a central module where they will be stored and may be consulted. The system is based on Arduino; this allows one program from the same development environment (IDE) of Arduino, simplifying the programming and obtaining greater accessibility to the users of the home. | 27% |

From the comparisons made, it is possible to say that the HEMSs represent a viable alternative for energy saving, since the reduction in consumption varies between 5.8 and 52.77%. The architecture of these HEMSs is constituted in a similar way, since in all of them, elements are defined that fulfil the functions of monitoring and control—central processing, a gateway, and a communication network that interconnects the household devices—the vast majority using low-energy consumption devices. Something important to mention is also that these jobs mostly seek to achieve a balance between savings and comfort; in addition, several of these works also make contributions to the use of renewable energy and the reduction in consumption at peak hours.

## 3. Materials and Methods

The methodology used for the development of the work was divided into 4 phases: the identification of environmental variables and the identification of devices for typical use in the home, the design and implementation of the WSN (wireless sensors network), the design and implementation of the management system and, finally, the performance evaluation of the system, which is the phase that yields the results of the work.

The evaluation of the system was carried out by measuring the power consumption in each device before implementing the system. The variables of current, voltage, power and energy consumption were measured. The data collected were used as a reference point in the evaluation of the results reported when measuring the same variables after the system was installed. In other words, the evaluation was based on the energy consumption data of some typical household appliances, before and after the system was installed. Then, it was possible to observe to what extent energy consumption was reduced.

According to the data provided by the bibliographic review and the previous surveys carried out, the typical household devices that are part of the test bench were selected. In this test bench with typical household loads, the offline energy consumption measurement was initially performed, and then the online consumption measurement was performed, and a comparison was made with the two results obtained. That is, for the offline measurement, the test bench was connected to the HEMS but only as an instrument to perform the measurements, and for the online measurement, the test bench was connected to the energy management system with all its functions active, and then with these results, the analysis was performed. This evaluation was carried out for 3 months.

## 4. Need Analysis in a Colombian Community

Before proposing the design of a low-cost architecture, a study was carried out that allowed us to analyze the consumption of electricity in the city of Sincelejo, for which a total of 1,066 surveys were applied, applying the technique of direct interview to families in each of the six socioeconomic strata in which the residential properties in Colombia are classified. According to this, it was possible to characterize the consumption of electricity that varies according to the stratum. From the analysis of the surveys, it was possible to determine the average number of occupants per household in each stratum, which is approximately four occupants for strata 1 to 6, so it is possible to say that a typical family in the city of Sincelejo is constituted of 4.2 members; the average daily consumption per stratum was also obtained, which increases as the stratum increases, meaning that the lower strata have lower energy consumption than the upper strata. This is shown in Figure 1.

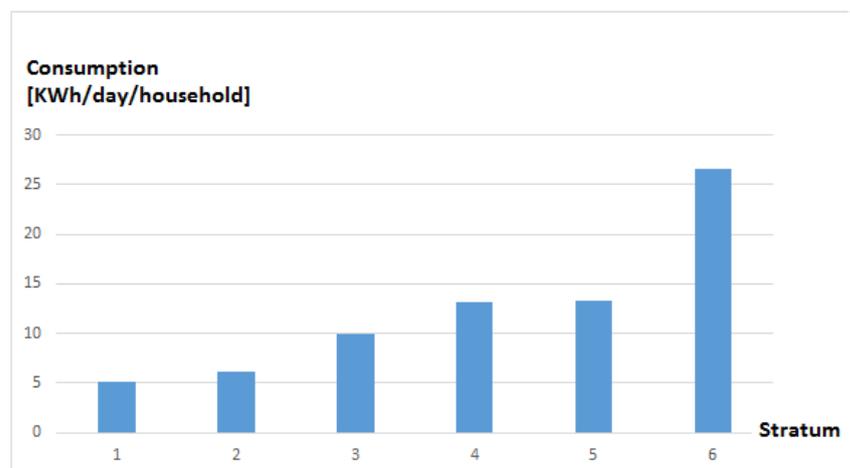

**Figure 1.** Daily consumption average per stratum.

## 5. Design of the Proposed Architecture

The architecture of the SAEH system is shown in Figure 2: the prototype is a small platform consisting of sensor–actuator nodes of four different types (temperature and humidity, luminosity, presence, and energetic consumption), a coordinating node and a monitoring and control center. Mobile devices can access the system through the internet using a gateway.

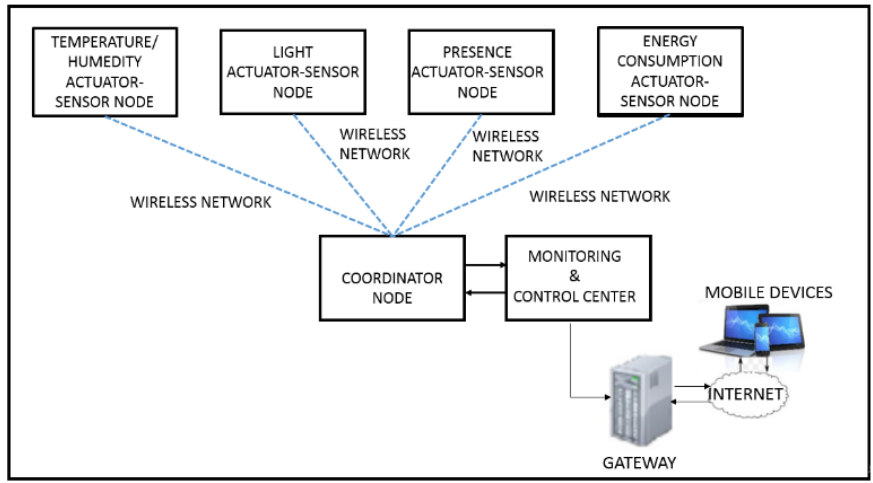

**Figure 2.** System Architecture.

### 5.1. Sensor–Actuator Nodes

The sensor–actuator nodes are the part of the system that are responsible for the monitoring and control of the devices. They capture the information and send it to the **coordinator node**, who then sends it to the monitoring and control center. The monitoring and control center are permanently listening to detect when the user has generated an event and send the command to the corresponding sensor–actuator node. In this way, it is possible to turn on or off the devices connected to the system.

Basically, these nodes have a processing unit (a microcontroller), a unit for wireless transmission, sensors, actuators, and a power supply block. Figure 3 shows the schema of the developed sensor–actuator node.

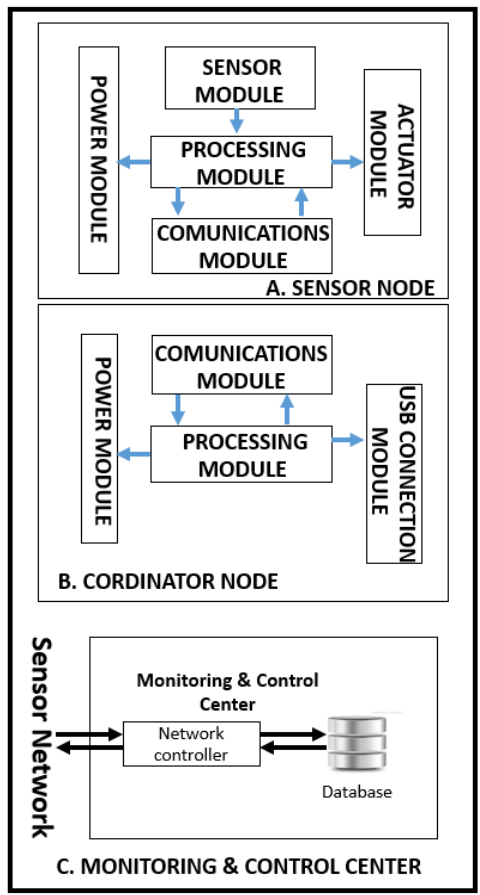

**Figure 3.** Scheme of the developed nodes and monitoring and control center.

Depending on the type of sensor–actuator node, the sensor fulfils a specific function as detailed below:

- The *Temperature and Humidity Sensor* allows the system to find the ambient temperature and relative humidity of the surrounding environment. The use of this sensor allows adjustments to the temperature of the room through the system.

- The *Light Sensor* determines the level of illumination of a room. In this way, it will be possible to increase or decrease the luminous intensity depending on the requirements of the user.

- The *Presence Sensor* is a sensor capable of detecting movement within its range of action. The sensor used detects levels of infrared radiation emitted by bodies with heat.

- The *Energy Consumption Sensor* can measure the energy consumption of an appliance. It measures the power consumed by current and voltage sensors. The proposed sensor measures five parameters: current, voltage, power, power factor and energy consumption in kWh. This sensor is basically a box connected to a power source to which the user connects the home appliance.

- Finally, the *Actuators* in the HEMS have the function of changing the state of a device after an event generated by the monitoring and control center from the reading of a sensor.

### 5.2. Coordinator Node

The coordinator node is responsible for collecting the data, receiving them through a radio interface and sending them through a serial port to the monitoring and control center. This node has a processing module, a module for wireless transmission, a module for serial communication and a power module. Figure 3b shows the outline of the coordinator node developed.

### 5.3. Wireless Transmission

The wireless sensor network (WSN) basically consists of a set of autonomous devices called nodes. The network has sensor nodes that use power sensors developed by us to capture information from the environment where they are located, and a coordinating node that is responsible for collecting the data, receiving them through a radio interface and sending them through a USB port. In the case of this project, a star topology was selected; that is, the sensor nodes communicate directly with the coordinator node and do not communicate with each other. The structure is the one shown in Figure 2.

### 5.4. Monitoring and Control Center

The monitoring and management center is in charge of receiving data from the wireless sensor network and performing network control by acting as a manager, storage and classifier of information, interface to the user and sending data to the Internet. In addition, the monitoring and control center is in charge of organizing the information received from the sensor nodes in a database, so that this information can be accessed by the user through an interface. For this, the monitoring and control center uses a storage unit, to which the operating system is installed. To access the system, mobile devices are connected to the Internet using a gateway. Figure 3c shows the diagram of the monitoring and control center.

### 5.5. Gateway

The gateway is a device that allows communication between the power management system and mobile devices. This consists of a processing module, a communication module, and a power supply.

*5.6. Mobile Devices*

    These are used to carry out the user's actions.

## 6. Technology and Implementation Details

    This section shows the technologies selected for the implementation of the system, Figure 4 shows the architecture with the technological details and project implementation.

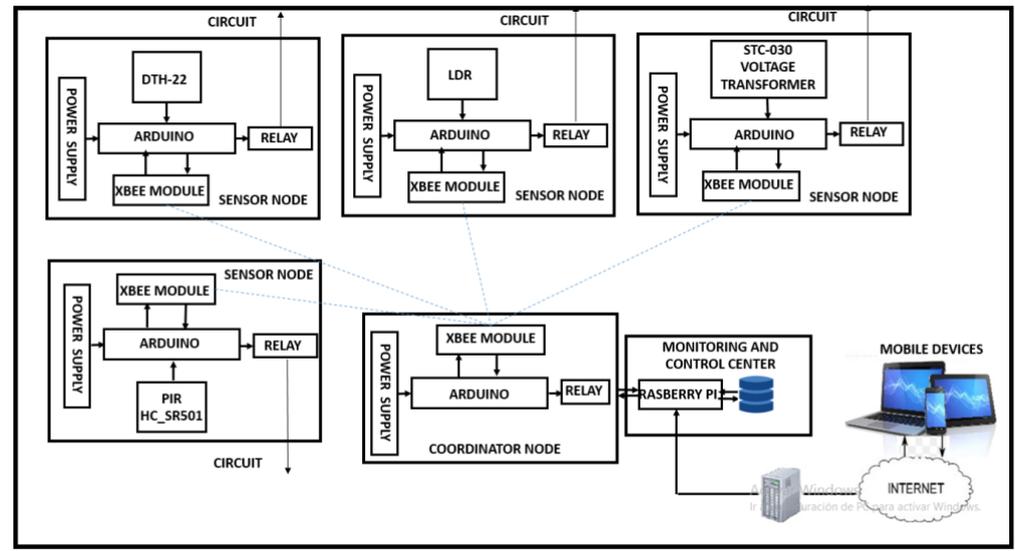

**Figure 4.** Complete System Architecture.

*6.1. Hardware Implementation*

    Before selecting the hardware to be used, a preliminary study was necessary to evaluate aspects such as robustness, processing and storage capacity, easy access to information in real time, energy consumption and ease of coupling the device with technologies provided by other manufacturers. Considering all the options available on the market, it was decided to use Arduino, an open electronic platform for prototyping based on flexible and easy-to-use software and hardware ([13,14]). Arduino is an alternative currently used for the development of WSN prototypes because it is an open-source technology, its programming language (processing) is simple and its hardware is flexible.

    Arduino can capture information from the surrounding environment through a wide variety of sensors through its input pins and can act on it by controlling a good number of actuators through its output pins. Plates can be built or purchased from a factory, and the software can be downloaded for free, all of which allows it to be tailored according to project requirements. Its development platform can be used with any operating system, and the programming language supports C ++ libraries. Additionally, Arduino is based on the ATMEGA168, ATMEGA328 and ATMEGA1280 microcontrollers, and the card is easily coupled with other technologies [15].

    After reviewing the technologies used for the transmission of data in WSN, it was decided to use Zigbee. This technology, developed by Zigbee Alliance, is one of the most used in wireless sensors networks. It is an open standard designed exclusively for data communication. Its main features are low cost, low power consumption, simplicity, high reliability, ease of implementation, compatibility, low latency and high capacity ([16,17]). Table 2 summarizes the main features of Zigbee.

**Table 2.** Zigbee Characteristics.

| Frequency of Operation | 868 MHz, 915 MHz, 2.4 GHz | Transmit Capacity | 23/250 Kbps |
|---|---|---|---|
| Range coverage | Until 3.2 Km | Potency of transmission (28 Kbps y 250 Kbps) | 1 mw, 10 mw, 100 mw |

For its implementation, the XBee modules, which are radio modules manufactured by DIGI international, are used: they have a wide variety of models within which one can choose the one that fits the needs of the system [18].

The Zigbee network consists of four sensor nodes and a coordinating node that is also mounted on an Arduino board.

For the prototype, the XBee radio series 2 was selected, which provides wireless endpoint connectivity to devices and the XBee shield to connect it to Arduino. This radius allows the communication of the sensor–actuator nodes with the coordinating node by sending and receiving information from and to the sensor–actuator nodes. XBee radios are programmed through the X-CTU (terminal program developed by Digi International).

The actuator–sensor nodes used DTH-22 as a sensor for the temperature and humidity node, an LDR for luminosity, PIR HC_SR501 for the presence of the user and for the power node, we used the voltage transformer and a current sensor, STC-030. Figure 5 shows the assembly of each of the sensors on the Arduino.

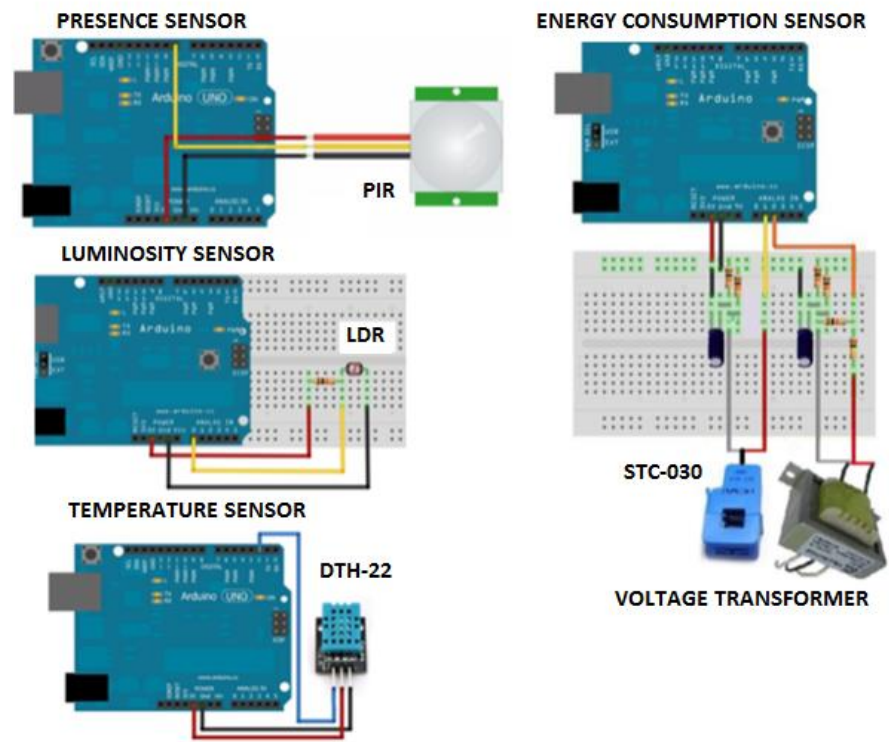

**Figure 5.** Assembly of each of the sensors to the Arduino.

For each of the different types of nodes, the XBee Shield expansion boards are directly connected to the Arduino UNO board by occupying determined pins of the Arduino board, where they take voltage to operate. The XBee Shield plate uses pins 0 and 1, which provide communication between the XBee module and the Arduino UNO board processor. Each of the sensor–actuator nodes have the necessary elements to measure the corresponding variable (energy, luminosity, presence and temperature and humidity) and receive the control signals necessary to manage the energy savings.

The energy consumption node is located in each one of the home appliances to be monitored. This node detects the current when it passes through the current sensor. A voltage signal proportional to that current is produced and delivered to the conditioning circuit; this circuit converts this signal into a form suitable for the input of the Arduino board. The voltage is detected through the voltage transformer, which reduces the value of the voltage and delivery to the conditioning circuit, which is also processed for the input of the Arduino board. The microcontroller of the Arduino board receives the values of current and voltage detected and is responsible for calculations of power, power factor and energy consumption in kWh. These values are sent to the XBee input to be processed and sent wirelessly to the coordinating node.

In the case of the presence, luminosity and temperature and humidity nodes, the detected signal is transformed into an electrical signal that is detected by the input of the Arduino plate; this information is processed by the microcontroller of the plate and sent to the input of the XBee for transmission to the coordinating node.

As actuator devices, we have chosen to use an electromechanical relay with a coil of 5 volts that works with the power of our microcontroller; despite operating at a relatively low voltage, we can control the switching on and off loads with a capacity up to 10 amps at a voltage of 125 Vac and 10 amps at a voltage of 30 Vdc. Changing the status of a household device is carried out through the relay module, which functions as a switch controlled by an electric circuit in which, by means of a coil and an electromagnet, a set of one or more contacts is operated, which allows the opening or closing of other independent electrical circuits. The monitoring and control center is permanently listening to detect when the user has generated an event and to send the command to the corresponding sensor–actuator node. In this way, it is possible to turn on or off the devices connected to the system, such as a bulb or a fan. In the case of a fan, depending on the temperature, the system could decide to turn on a fan or not, or leave the decision in the hands of the user. In the case of a light bulb, if the luminosity decreases from a certain point, then the system could light a lamp, and if it increases, it could either turn it off or finally allow the user to decide.

The monitoring center can be a computer with any operative system; however, it has chosen to use the platform Raspberry pi for its small size and affordable price. Raspberry pi, developed by the Raspberry Pi Foundation [19], handles an ARM processor, 512 mb of RAM and a VideoCore graphics processor; for information storage, it uses an SD card. It has as outputs USB ports and a 10/100 Ethernet controller. This monitoring center will be connected to the coordinating node Zigbee via USB.

The gateway consists of an Arduino UNO, Ethernet Shield and a power supply. The Shield Ethernet is the one that allows us access to the local network that we have. By means of an Ethernet cable with RJ45 at both ends, the gateway is connected to the device (router or switch) that provides Internet access to the home. Figure 4 shows how the above-mentioned components are connected.

### 6.2. Software Implementation

For the development of the project, we chose the platform Arduino UNO, which supports the management of the sensors. The programming of the sensors was undertaken by connecting these to the Arduino board (some of them required a coupling circuit), and the Arduino was connected to a computer using a USB interface, to be programmed by means of the IDE (integrated development environment) of Arduino. UNO loads and compiles the code (script) to the Arduino microcontroller. The codes programmed in the sensor nodes allow the reading of the sensors, the control of the actuators and the configuration of the Zigbee connection.

The X-CTU software is used to program the Zigbee to set parameters for wireless network operation. In order to program the XBee modules, it was necessary to use a programmer that connected to the XBee module and to a computer using a USB interface to perform programming using XCTU.

The control center application runs on the Raspbian operating system for Raspberry Pi, a derivative of Debian. When an Arduino is connected via USB to Raspberry Pi, the Arduino coordinator of the network forwards via USB everything received from the ZigBee network. All data forwarded by the coordinator can be read in the control center as if it were a current file.

## 7. Testing

It was possible to build the prototype of an HEMS, a low-cost electricity management platform that that operates in three stages:

- Data collection: Data are received; that is, the status of all sensors.
- Processing: The status of all the sensors is displayed in the monitoring and control center. Then, the data are available to the user through the user application.
- Control: As the user can know the status in real time of the devices connected to the system, they can perform control actions on these from the user interface installed on their mobile device.

The acquisition of data was made in the Electronica laboratory of the University of Sucre over five days, in which different samples of each variable were taken in different ways, depending on the variable, which allowed the observation of their behavior. The results obtained from the comparison between the sensor nodes of the prototype and the standard measurement instruments show that the accuracy of the proposed wireless data acquisition circuits, using the Zigbee module, is considered satisfactory, as evidenced by the functional tests carried out, which are detailed later in this section. The implementation of the wireless network in the star topology in the developed system is simple and transparent to the user due to the support provided by the XBee communication module (Zigbee). In addition, this topology has proved to be satisfactory for the application since it allows the coordinating node to obtain the data collected by the sensor nodes. The data are also received without alterations in the monitoring and control center, and then they are correctly registered and stored in the system database. The gateway used allows the data to be displayed correctly by the user in the mobile application, and the user can control the devices in the HEMS.

Tests of functionality, connection to the network and data transmission were carried out to verify the functioning of the HEMS.

### 7.1. Functionality Tests

Functional tests were designed for each of the elements that are part of the system. First, tests of the cards corresponding to each sensor node were carried out. The performance of the prototype cards was tested in terms of data transmission speed and the accuracy of the data received. To test the accuracy of the received data, standard measuring instruments are used. For the case of the energy consumption node, a digital multimeter was used to determine the current and voltage readings of some typical household appliances. The current and voltage signals received at the coordinator node were compared with the measurement of the digital multimeter. In the case of the temperature and humidity sensor, a data acquisition comparison was made between the sensor node of the prototype and a digital environmental hygrometer thermometer, taking a value for temperature and humidity every hour between 9:00 a.m. and 5:00 p.m. For the luminosity sensor, the results obtained by the Arduino board sensor were compared with the data thrown by a digital luxmeter; these data are taken with different luminous intensity in a room. The results of the test are recorded in Tables 3–5, respectively.

**Table 3.** Energy consumption sensor.

| Home Appliance | Arduino Board | Digital Multimeter |
|---|---|---|

|  | Current | Voltage | Current | Voltage |
|---|---|---|---|---|
| Blender | 3.18 A | 127 V | 3.20 A | 126.9 V |
| Fridge | 4.52 A | 118 V | 4.55 A | 118.3 V |
| Computer | 1.18 A | 115 V | 1.20 A | 114.6 V |
| Fan | 1.36 A | 114 V | 0.6 A | 114.5 V |
| T.V. | 5.28 A | 127 V | 1.34 A | 126.4 V |
| Washing machine | 15.28 A | 110 V | 15.40 A | 110.5 |
| Air conditioner | 10.95 A | 219 V | 11.01 A | 218.9 V |

**Table 4.** Temperature and humidity sensor.

| Hour | Arduino Board | | Digital Environmental Hygrometer Thermometer | |
|---|---|---|---|---|
|  | Temperature | Humidity | Temperature | Humidity |
| 9 | 30.57 °C | 40.00% | 30.8 °C | 39% |
| 10 | 31.54 °C | 39.90% | 31.9 °C | 39% |
| 11 | 32.52 °C | 38.30% | 33 °C | 38% |
| 12 | 33.50 °C | 37.50% | 33.4 °C | 38% |
| 13 | 34.47 °C | 36.40% | 34.2 °C | 36% |
| 14 | 34.90 °C | 36.10% | 34.8 °C | 36% |
| 15 | 33.01 °C | 37.50% | 33.1 °C | 37% |
| 16 | 31.46 °C | 39.00% | 31.6 °C | 39% |
| 17 | 30.50 °C | 39.50% | 30.4 °C | 40% |

**Table 5.** Luminosity Sensor.

| Luminous Intensity | Light Sensor Arduino Board | Digital Luxometro |
|---|---|---|
| Very bright room | 330 lux | 325 lux |
| Room with medium lighting | 155 lux | 150 lux |
| Room with dim light | 82 lux | 79 lux |
| Dark room | 0.09 lux | 0 lux |

As can be seen in Tables 3–5, the values obtained with the Arduino plates designed for the project are reliable, since the values are very close to those thrown by commercial measuring instruments. This can be verified by observing the standard deviation values for each parameter reported in Table 6, where the standard deviation of the measured data with the multimeter and the data measured with the Arduino are compared. From there, it is possible to deduce that there is not much difference between both cases, which suggests that the values thrown by the Arduino node are reliable.

**Table 6.** Standard deviation.

| Parameter | Arduino | Commercial Instrument |
|---|---|---|
| Current | 5.26046847 | 5.688621475 |
| Voltage | 38.5289007 | 38.43308966 |
| Temperature | 1.60582533 | 1.591383046 |
| Humidity | 1.46125897 | 1.414213562 |
| Luminous | 140.536165 | 138.6085615 |

In the case of the presence sensor, the tests performed were different. Table 7 shows the different values for the tests in distance and degrees of the sensor with respect to the individual and the detection or not of presence of the Arduino board. To measure the speed of data transmission in the cards, the sampling time was taken into account, which is the time elapsed between two consecutive measurements. According to tests carried out using a stopwatch and at a small distance, at which there was no loss of packets (1 m), the cards sent one packet per second, as had been programmed in the microcontroller for the purpose of testing.

**Table 7.** Presence Sensor.

| Distance (m) | Degrees | Detection |
|:---:|:---:|:---:|
| 1 m | 0° | Yes |
| 1 m | 25° | Yes |
| 1 m | 50° | No |
| 2 m | 0° | Yes |
| 2 m | 25° | Yes |
| 2 m | 50° | No |
| 3 m | 0° | Yes |
| 3 m | 25° | Yes |
| 3 m | 50° | No |
| 4 m | 0° | Yes |
| 4 m | 25° | Yes |
| 4 m | 50° | No |
| 5 m | 0° | Yes |
| 5 m | 25° | Yes |
| 5 m | 50° | No |
| 6 m | 0° | No |
| 6 m | 25° | No |
| 6 m | 50° | No |

*7.2. Test Connection to Zigbee Network*

In this section, we will see the tests performed to test the extent of communications between transmitter and receiver of the XBee modules, checking if the nodes work and if their sensors collect data correctly. We moved the sending node through different points of the proof area, checking its scope and study where the coordinator stopped receiving the data correctly. Table 8 shows the different distances at which the scope of the communication was checked.

**Table 8.** Zigbee test connection.

| DISTANCE (m) Between Tx and RX | INTENSITY OF THE RECEIVED SIGNAL |
|:---:|:---:|
| 5 m | −39 |
| 10 m | −53 |
| 15 m | −69 |
| 20 m | −80 |

With the tests, it was found that with increasing distance, the intensity of the signal decreased. With the tests, it was proved that from 16 m, the transmission presented packet losses due to the obstacles present in the house such as walls and furniture; however, until 15 m, the transmission did not present packet losses. Although the Xbee series 2 does not comply with the distance of 30 m specified by the manufacturer, for our case, it was useful, since this distance is between the typical dimensions of residential buildings in the city of Sincelejo. Another important factor after the tests was the position of the receiver: the higher the elevation with respect to transmitter, the greater the scope.

*7.3. Evidence of Sending Data*

The results of this phase showed that even if both end node stations sent the sensor reading to the base station at the same time, no interference occurred. The data received from each station (Node 1, 2, 3 or 4) are displayed correctly. For the realization of this test, there were five nodes, and the test was carried out at the University of Sucre. The calculation of the distances to which the nodes could communicate was made beforehand, and then the location of these was determined. After this, the communication between the nodes was established, for which the configuration of a coordinating node and four nodes as final devices was previously made. In this way, the network was established, and the arrival of information was observed correctly to the XCTU program installed on a PC where the coordinator node was connected. In the previous section, the transmission distances are detailed.

We also developed data accuracy tests and communication between the coordinator and the monitoring and control center (raspberry pi) being able to observe that it has a very stable operation thanks to the USB interface, which makes communication more secure, accessible, easy and fast, which allows the monitoring and control center to obtain temperature and humidity data, luminosity, presence and energy consumption collected by the sensor nodes.

## 8. Validation of the Systems

The prototype of the system designed and developed allowed us to measure the energy consumption of household appliances, as well as other variables through a network of wireless sensors. It also allowed end users to access this information in real time and control the devices in the network. The results showed that the proposed system is efficient and accurate in the measurement of the variables. The prototype developed with the Arduino UNO and Zigbee microcontroller is low cost, and the proposed mobile application for the control of the system allows a gamification approach to be developed in order to influence the behavior of users regarding the reduction in energy consumption in the home.

For the evaluation of the energy management system, a user type stratum 3 was defined, for which a weekly load profile corresponding to the electrical appliances of the test bench that is implemented in the HEMS is proposed. This would be performed by comparing the traditional profile and the profile obtained after the implementation of the HEMS, and by determining the impact caused by HEM on energy consumption and the saving of electrical energy achieved through its use.

**Traditional weekly upload profile for a user:** The construction of the weekly load profile is the result of the review of two main components: first, the electrical appliances that will make up the test bench, which have been selected according to the most relevant for stratum 3, and are obtained from the characterization of electrical energy consumption in the residential sector of Sincelejo Colombia [5], and secondly, the definition of the typical user and their activities and patterns of use of electrical energy, taking into account the electrical appliances included in the test bench and the measurements made during the day. Finally, the weekly load profile is the combination of the different behaviors of the user during the week; the user profile represents the household with all its members.

The characterization of electrical appliances and devices: The electrical appliances listed in Table 9 have been selected for the test bench, which were used for the construction of the load profile. The main characteristics are described for each appliance.

**Table 9.** Characteristics of the electrical appliances of the test bench.

| Appliance | Manufacturer | Model | Power [W] |
|---|---|---|---|
| TV | Samsung | LCD 450 | 54 |
| Fan | Samurai | Max Air FS | 52 |
| Air-conditioning | Panasonic | CS-YS12TKV | 3520 |
| Personal computer | Dell | Vostro 260s | 250 |

| Incandescent light bulb | Phillips | E27-A55 | 72 |

**Definition of the typical user:** This definition is made in order to show the results of the HEMS implementation in a practical and specific environment. The characteristics of the typical user for our case is a stratum 3 household made up of a family of four, with a father, mother and two children

**Definition of the functionality of the system:** The description of the functionality is made through Table 10. For each listed functionality, a brief description is given of how the HEMS operates and the respective management component is established, which becomes a management tool that allows the execution of a management action energy later in the application of the HEMS to the type use.

**Table 10.** Description of system functionality.

| Functions | How the HEMS does it | Management Tool |
|-----------|---------------------|-----------------|
| Disaggregated consumption in real time | The energy consumption sensor–actuator nodes measure the consumption of the household appliance to which they are connected, and these data are sent to the coordinating node, which sends it to the monitoring center where it is processed to be presented to the user. | Disaggregated consumption monitoring |
| Information Processing | The monitoring and control center will store the data in a MYSQL database; these data are available for the user to view through the website and the mobile application. Information processing is performed using a Rasberry Pi. | Storage and access to consumption data, execution of actions on the system. |
| Remote Control ON/OFF | The user, through a mobile device, sends control actions to the monitoring and control center, who, in turn, sends these to the coordinator, to finally be sent to the sensor–actuator nodes, which have ON/OFF control. | Control of household appliance consumption. |

Usage profile: Figure 6 records the daily usage profile for each appliance connected to the test bench. This profile allows us to easily identify the start and end times of the use of each appliance and therefore the duration of its use; as it can be seen, the usage time differs depending on the appliance.

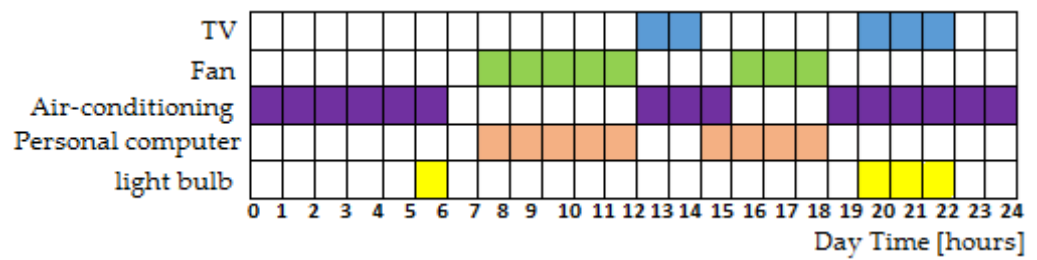

**Figure 6.** Use profile per appliance.

For the validation of the system, the test bench was installed in a single room. When the system is tested in a real home, these profiles will be built with the information collected every 24 h for n days to predict which devices will be used, at what hours and for how long; data that will be used to feed the game.

Comparison of the Consumption profile per appliance before and after the implementation of the HEMS: This comparison between the consumption profiles per appliance seeks to highlight the benefits of its implementation in terms of reducing energy consumption in kWh, this comparison is shown in Table 11.

**Table 11.** Comparison of the Consumption profile by appliance.

| Comparison | Light Bulb | Fan | Computer | TV | Air Conditioning | Weekly Total |
|---|---|---|---|---|---|---|
| Weekly profile without HEMS kWh/Week | 1.5 | 2.9 | 1.07 | 28.05 | 273.22 | 306.74 |
| Weekly profile with HEMS kWh/week | 1.17 | 2.7 | 1.0165 | 25.245 | 193.98 | 224.12 |
| Consumption reduction in kWh/week | 0.33 | 0.203 | 0.0535 | 2.805 | 79.234 | 82.62 |
| Percent decrease | 22% | 7% | 5% | 10% | 29% | 27% |

Table 11 represents the contrast between the weekly load profile of the typical user without HEMS and the same load profile after the implementation of the HEMS. Achieving with this implementation a total reduction in consumption of 82.62 kWh/week, equivalent to a 27% weekly decrease in energy consumed.

The design criteria have focused on ensuring that the system is affordable, reliable, easy to use and adaptable to the needs of the owners. To accurately manage these functions, it is crucial to continuously monitor a number of parameters such as temperature, humidity, the presence of people in the room and the energy consumption of each device connected to the test bench. The system shows the total energy consumption and the disaggregated consumption for each appliance, which allows the users of the system to take measures to reduce energy consumption through manual control through switches or through the internet with the use of mobile devices.

## 9. Results

Due to the need to save energy in the home, HEMSs have been developed; these systems have a basic architecture, and they are supported by HAN for the transfer of information. This paper presents a design of an HEMS that aims to help the user to participate in the optimization of energy consumption. Additionally, the research based on a needs analysis previously conducted by the authors includes the development of a system that guarantees low acquisition and start-up costs, ease of implementation, compatibility with existing communication systems, hardware and free software and ease of expansion.

The graph in Figure 7 shows us the weekly energy consumption for the offline and online system per appliance; the reduction in consumption can be clearly seen when using the HEMS. Consumption is represented in kW/h on the vertical axis, and the appliances are arranged on the horizontal axis.

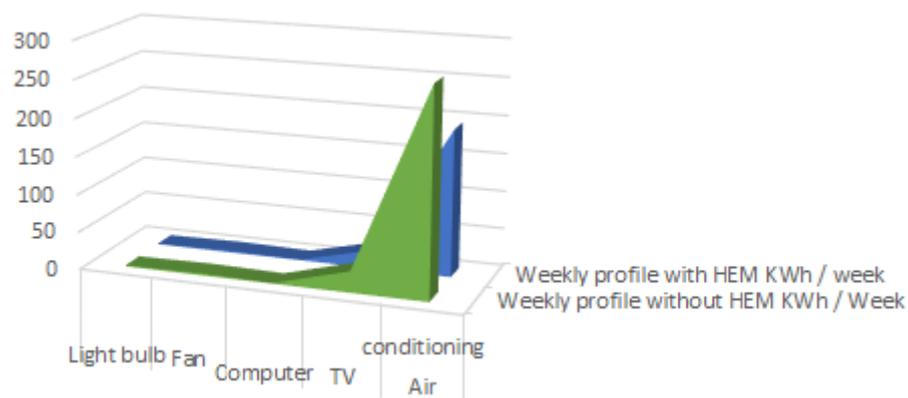

**Figure 7.** Comparison of weekly consumption by appliance.

The graph in Figure 8 reflects a reduction of 27% in consumption when the HEMS is used. Greater savings are expected when the system is complete.

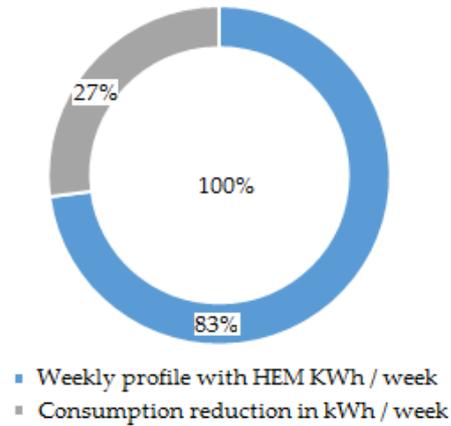

Weekly profile without HEM KWh / Week

27%

100%

83%

■ Weekly profile with HEM KWh / week
■ Consumption reduction in kWh / week

**Figure 8. Impact of HEM usage in energy consumption**

In Colombia, commercial home automation systems remain unaffordable for the vast majority of lower- and middle-class families. The use of cheap microcontrollers such as Arduino has allowed the implementation of smart systems at a much lower cost with a large number of features like those offered by commercial systems. Table 12 shows a comparison of the proposed system with other commercial systems that can be purchased in Colombia; this shows the cost and its main characteristics.

**Table 12.** Comparison of the proposed HEMS with commercial systems.

| System | Characteristics | Price (Colombian pesos) |
| --- | --- | --- |
| Energy resources solutions [14] | It integrates base load changes from energy efficiency measures, combining them with contextual information such as real-time network combination, temperature data, among others. | 6,551,522.92 |
| Green Energy [20] | The system shows you current energy usage and has a budget function to help householders keep track of costs. | 1,806,720.00 |
| Home Automation [21] | Provides automated lights, locks, thermostat, smart speaker, hub, and smart plugs | 3,013,440.00 |
| Smart Wi-Fi thermostat [15] | Automatically adjusts your thermostat using your smartphone's location to easily save energy. | 1,796,010.24 |

| Savant system [22] | You can control lighting, weather, entertainment and security and save energy by configuring your system to turn off lights in empty rooms from a single app. | 4,818,490.56 |
| Proposed System | The system integrates a light intensity sensor, a temperature and humidity sensor, a motion sensor and an energy consumption meter. The data acquired by these sensors are displayed on an LCD screen built into each meter to provide information to the user on the status of the observed parameters. | 1,200.000 |

As can be clearly seen in Table 12, the proposed HEMS is the one with the lowest cost, thanks to the use of Arduino boards that are cheap compared to other microcontroller platforms and reduce the cost of the system. In addition, Arduino has a low power consumption and its open-source software platform makes it easy to use, which makes it a useful tool as an initial prototype in home control systems. These reasons allow it to be widely used for control in electronic systems.

## 10. Conclusions and Future Work

This work proposes a low-cost HEMS model, since cheap devices, as well as free software, have been used in its design. Its lower commercial value, when compared to other alternatives in the Colombian market, allows its deployment in a greater number of homes. In addition, the proposed HEMS has been evaluated in a pilot by taking consumption data before and after its deployment; the pilot reveals a 27% saving in energy consumption. To obtain greater savings, user interaction with the HEMS can be carried out through a mobile application, given that a large part of the population have smart phones [23–26]. The monitoring and control center is responsible for communications between the HEMS and the user. That is to say, it is the point of union between the monitoring and control center, the database and the mobile application. It also produces notifications and the exchange of data between the hardware and the mobile application. These interactions will be carried out through the local server that collects the information from the sensors and waits for the possible actions that the user makes from the control. The mobile application performs data management to allow the user to have the values of the variables updated in real time, as well as historical data, graphs, and notifications of events. This application will allow to personalize the process, because each home is different, in addition it is thought to use a gamification scheme to improve the efficiency of the HEMS, developing an application that turns into a gamified control of household appliances in the home.

Because the HEMS can show not only the total accumulated consumption, but also disaggregated consumption, that is, the accumulated consumption in each appliance, a user can find out the daily, weekly and monthly consumption of each appliance. You can also identify which device consumes more power. With the help of energy consumption information, a user can reduce energy consumption by changing habits that cause energy waste. To achieve this effect on users, researchers are working on the development of a gamification scheme to improve the efficiency of HEMSs, to make this task entertaining for the user. The proposed gamification architecture consists of a power management system, an activity management server, and a gamification engine. The generation of gamified tasks that are sent to the user to persuade and improve their consumption habits and subsequently obtain energy savings are based on the information provided by the predictive model that will be added to the system. The data generated in this stage will allow, together with the HEMS information, model to be re-trained to adjust the precision of the prediction. Once the model learns, personalized gamified activities can be proposed based on the user's consumption patterns.

The validation tests carried out were only to check the correct functioning of the prototype before its implementation. The results section includes the results obtained by comparing the load profiles before and after the installation of the HEMS for a home; this is to check the viability of the HEMS because the project is in progress. To validate the system in its final phase, the prototype will be installed in five homes, and the average energy consumption of household appliances will be measured for a year, 6 months with the HEMS offline and 6 months with the HEMS online. Then, calculations will be made to obtain the percentages of energy saved.

**Acknowledgments:** The authors thank SENA and the Sucre University for their partial support of this work through the SAEH project. "This work is also partially funded by the European Regional Development Fund (ERDF) and the Government of the Autonomous Community of Galicia under the agreement with the Center for Research in Information and Communication Technologies (AtlantTIC) and the Ministry of Economy and Competitiveness within the framework of the National Research Plan (project TEC2014-54335-C4-3-R and TEC2017-84197-C4-2-R).